\documentclass[12pt]{article}
\usepackage{amsfonts,latexsym,amsmath,epsf,latexsym,amssymb}
 
\newtheorem{postulate}{Postulate}

\newtheorem{theorem}{Theorem}

\def\N{{\mathbb{N}}}

\newcommand{\cO}{{\cal O}}

\newcommand{\proof}[1]{{\bf Proof.} #1 \hfill $\Box$\vspace{0.5cm}}

\newcommand{\ket}[1]{|#1\rangle}
\newcommand{\bra}[1]{\langle #1|}

\title{Measuring  $4$-local $n$-qubit observables 
could  probabilistically solve PSPACE}

\author{Pawel Wocjan\thanks{e-mail: {\protect\tt
\{wocjan,janzing,decker\}@ira.uka.de}}\,, Dominik Janzing, 
Thomas Decker, \\and Thomas Beth\\ \small
Institut f{\"u}r Algorithmen und Kognitive Systeme, Universit{\"a}t
Karlsruhe,\\[-1ex] \small Am Fasanengarten 5, D-76\,131 Karlsruhe,
Germany}

\date{August 1, 2003}

\begin{document}

\maketitle

\abstract{We consider a hypothetical apparatus that implements
measurements for arbitrary $4$-local quantum observables $A$ on $n$
qubits. The apparatus implements the ``measurement algorithm'' after
receiving a classical description of $A$. We show that a few precise
measurements, applied to a basis state would provide a probabilistic
solution of PSPACE problems. The error probability decreases
exponentially with the number of runs if the measurement accuracy is
of the order of the spectral gaps of $A$.

Moreover, every decision problem which can be solved on a quantum
computer in $T$ time steps can be encoded into a $4$-local observable
such that the solution requires only measurements of accuracy
$O(1/T)$.

Provided that BQP$\neq$PSPACE, our result shows that efficient
algorithms for precise measurements of general $4$-local observables
cannot exist. We conjecture that the class of physically existing
interactions is large enough to allow the conclusion that precise
energy measurements for general many-particle systems require control
algorithms with high complexity.}

\section{Measuring $k$-local observables}
A characteristic feature of quantum theory is that there exists an
abundance of mutually incompatible observables (described by
self-adjoint operators $A$) for every quantum system and it is by no
means obvious how to implement measurement procedures for all these
quantities. On a quantum computer one could in principle measure every
observable $A$ as follows: Find a unitary transformation $U$ which
diagonalizes $A$ with respect to the computational basis. Then $A$ is
measured by implementing $U$ and measuring the logical state of each
qubit. By identifying each binary word with the corresponding
eigenvalue of $A$ this procedure reproduces all probabilities
\[
p_j := tr(\rho P_j)
\]
correctly where $\rho$ is the density matrix of the quantum register
and $(P_j)$ is the family of spectral projections of $A$.

However, the implementation of the diagonalizing operation $U$ will in
general be hard. Therefore, one may restrict the attention to specific
classes of observables. It is natural to consider observables with
physical relevance. For example, the quantum observable ``energy'',
mathematically described by the self-adjoint operator $H$ (the
Hamiltonian), is certainly one of the most important observables in
physics.  It determines the dynamical and thermodynamic behavior of
the considered quantum system. Furthermore the eigenstates of the
Hamiltonian, the energy levels, are ``directly'' observable in many
physical situations. For instance, in spectroscopy the eigenvalues of
the Hamiltonian determine the frequencies of emitted or absorbed
photons. Nevertheless, the determination of the energy levels in
interacting many-particle systems is in general a difficult task.

To explain this more explicitly, we need to describe the class of
operators which is considered. First we note that physical interaction
Hamiltonians usually satisfy some locality condition in the following
sense. We call an $n$-qubit operator $k$-local if it is a sum of
operators which act on at most $k$ particles non-trivially.  For
fundamental interactions between real physical particles one has more
specific statements and may restrict the attention to
pair-interactions. Nevertheless, $k$-local interactions among qubits
are physically reasonable. They may describe effective Hamiltonians
and there is not necessarily a one-to-one correspondence between
qubits and physical particles ($l$ qubits may, for instance, describe
the state of one particle). The following results show that it is in
general difficult to compute the spectrum of $k$-local Hamiltonians.

The problem of determining the lowest energy value of a (classical)
spin-spin interaction of Ising type is known to be NP-complete
\cite{Barahona:82,Pawelcompass}. For interacting qubits determing the
lowest energy value is even QMA-complete (``Quantum-NP'') if one
allows $3$-local interactions only \cite{KitaevShen,KempeRegev}. Note
that in these NP and Quantum-NP problems the task is not to determine
the lowest eigenvalues {\it with high precision}. The demanded
accuracy is only inverse polynomially in the number $n$ of interacting
qubits.  This has implications for the measurement procedure above:

The unitary $U$ that maps the eigenvectors of $A$ to the computational
basis states is only helpful for measuring $A$ if the correspondence
between computational basis states and the eigenvalues of $A$ is
known. Therefore, this method would require to know the spectrum of
$A$. For $2$-local or $3$-local observables one would need the
solution of NP- and QMA-hard problems, respectively.

In this paragraph we will explain that measurements of $k$-local
observables $A$ are possible up to inverse polynomial accuracy {\it
without} using any knowledge on the spectrum of $A$.

Here we do not need a precise definition of accuracy, we only demand
that the following condition is satisfied:

\begin{postulate}[Measurement accuracy]\label{def:measure}${}$\\
A measurement with accuracy $\Delta \lambda$ has the following property:
For all density matrices $\rho$ 
the probability to obtain an outcome in the interval
$I:=[\lambda_j-\Delta \lambda, \lambda_j +\Delta \lambda]$ is at least
$
(3/4)\,  {\rm tr} (\rho P_j)
$.
\end{postulate}
Our result is not sensitive to the particular
definition of accuracy. However, it is convenient to work with the 
formulation above.

Now we describe how to implement approximative measurements.  The idea
is that for every $k$-local $A$ (with $k$ constant) the corresponding
time evolution $U_t:=\exp(-iAt)$ (if $A$ is interpreted as a
Hamiltonian $H$ of a quantum system) can be simulated efficiently in
an approximative sense. Explicitly, it has been shown that the
simulation of $U_t$ with elementary gates up to an error of $\epsilon$
(with respect to the operator norm) requires $O(t^2/\epsilon)$ gates
\cite{Lloyd}. Now we can choose $t$ in such a way that there is a
one-to-one correspondence between the eigenvalues of $U_t$ and
$A$. This is the case whenever $\|A\| t\leq \pi$. An upper bound on
the norm of a $k$-local operator is easy to get. We assume without
loss of generality that each $k$-local term is upper bounded by the
value $1$. There are at most
\[
{n \choose k} < n^k
\]
$k$-local terms. Therefore, one has $\|A\|=O(n^k)$. For an appropriate
value of $t$ we can implement measurements of the ``observable'' $U_t$
using\footnote{Note that also a unitary operator defines in a
canonical way an observable by its spectral projections if one allows
complex measurement outcomes.}  the quantum state estimation procedure
\cite{ClevePhase}. We will briefly sketch the idea. In the following
we drop the index $t$.

\begin{figure}\label{energyfig}
\centerline{\epsfxsize1.0\textwidth\epsfbox{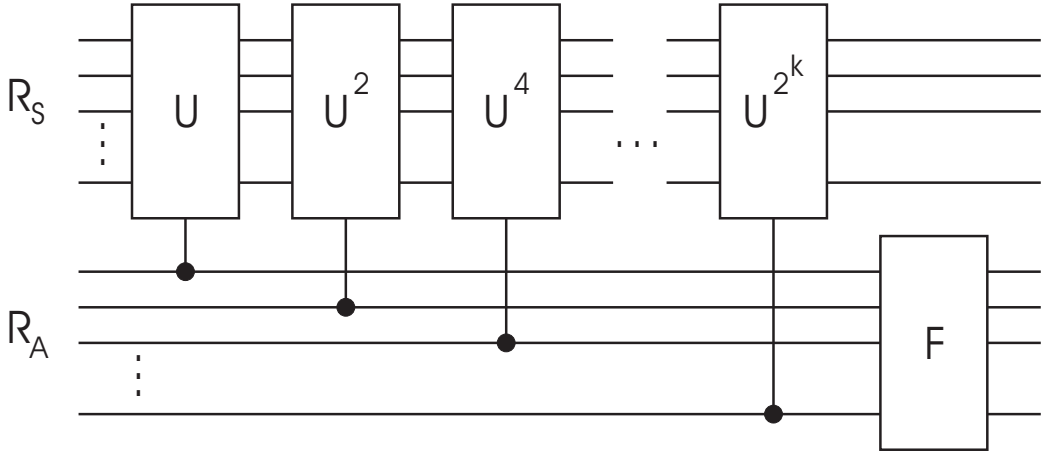}}
\caption{Circuit for performing measurements of an observable $A$. 
The powers of $U:=\exp(-iAt)$ are implemented as conditional gates
controlled by the ancilla register.}
\end{figure}

The circuit for phase estimation is shown in Fig.~\ref{energyfig}. It
acts on the registers $R_S$ and $R_A$. Controlled-$U^j$ gates are
implemented in such a way that $U^j$ is performed on $R_S$ if and only
if the ancilla register $R_A$ is in a state corresponding to the
binary word $j$. This can be done by implementing controlled
\[
U^{2^l}
\]
gates which are applied if and only if the $l$th qubit of the ancilla
register is in the state $|1\rangle$. The algorithm starts with an
equally weighted superposition
\[
\frac{1}{\sqrt{2^m}}\sum_{j=1}^{2^m}|j\rangle \,,
\]
of all ancilla register states where $m$ is the number of ancilla
qubits. After applying the controlled $U^j$-operation the discrete
Fourier transform of size $2^m$ is applied to the ancilla
register. Then the eigenvalues of $U$ can be readout with an error of
the order $1/2^m$. Now we consider the running time of this scheme
(depending on the accuracy). Obviously, this depends on the running
time for implementing the controlled-$U^j$ operations.  The natural
method to implement $U^j=\exp(-iAjt)$ is to simulate the time
evolution with Hamiltonian $A$ for the time $tj$. The substitution of
the corresponding gates by controlled gates is straightforward.
However, $j$ grows exponentially with $m$. Consequently, this method
requires exponential running time for exponential accuracy. It is
likely that all possible schemes for measuring $A$ precisely share
this disadvantage.

For a black-box unitary $U$ it is clear that exponential accuracy
requires exponential time since the black-box unitary $U$ has to be
applied an exponential number of times. This can be seen by subjecting
two state vectors to different unitaries $U$ and $\tilde{U}$ with the
same eigenstates but slightly different eigenvalues. Then $U^l$ and
$\tilde{U}^l$ can only lead to distinguishable states for large
$l$. In \cite{Jan01} it is described how to convert the time evolution
$\exp(-iHt)$ according to an unknown pair-interaction Hamiltonian $H$
to a controlled-$\exp(-iHt)$ evolution. This shows that black-box
settings for unknown $A$ do in principle make sense for energy
measurements.

For non-black box interactions we cannot obtain lower bounds on the
measurement complexity by similar arguments since the apparatus
receives a classical description of the observable to be measured.
However, the result of this paper suggests that even if the
interaction is known there is no efficient measurement scheme with
exponential accuracy. We show that measurements of $4$-local $n$-qubit
observables $A$ could be used to solve PSPACE-problems in polynomial
time provided that the accuracy is sufficient to distinguish between
the different eigenvalues of $A$.

One may ask whether there may be any physical processes for measuring
$4$-local observables that do not rely on quantum circuits (consisting
of elementary gates). For instance, one may guess that a measurement
of the energy of a system is simpler than a measurement of an
arbitrary $k$-local observable because energy is a specific observable
determining many physical aspects of the system. But note that the
quantum version of the Strong Church-Turing Thesis (compare
\cite{Shor,NC})  states that
every problem that can be solved efficiently using some physical
process can be solved efficiently by a quantum computer.

If there existed any efficient scheme for precise measurements of
$4$-local observables this would imply either of the following
statements:

\begin{enumerate}
\item The measurement process cannot efficiently be simulated on a
quantum computer (in contrast to the strong quantum Church-Turing
Thesis).
\item There are polynomial time algorithms to solve probabilistically
PSPACE problems, i.e., PSPACE $=$ BQP.
\end{enumerate}
Assuming that both implications are unlikely, our result strongly
suggests limitations for future quantum measurement technology.

The structure of the paper is as follows. In the next section we
consider a class of quantum circuits with polynomial size. They could
solve PSPACE problems if they were applied an exponential number of
times on a polynomial number $m$ of qubits.

In Section~3 we describe how to construct a $4$-local observable
corresponding to this circuit in such a way that precise measurements
would solve PSPACE problems in polynomial time.

\section{Characterizing PSPACE by circuits}
The complexity class PSPACE is usually defined with respect to the
Turing machine model \cite{HU:79}. PSPACE is the class of all
languages recognizable by polynomial space bounded deterministic
Turing machines that halt on all inputs \cite{GJ}.

For our purposes we need a characterization of PSPACE with respect to
quantum circuits. In particular, we need the result that every PSPACE
language can be recognized by applying an appropriate circuit many
times.

\begin{theorem}[PSPACE]\label{th:pspace}${}$\\
For every language $L$ in PSPACE there is a polynomial-time uniformly
generated family of quantum circuits $(V_l)_{l\in \N}$ consisting of
$s_l=poly(l)$ elementary quantum gates and acting on $m_l=poly(l)$
many qubits. The circuit $V_l$ decides whether an input string $x$ of
length $l$ is an element of $L$ in the following sense.

There is a 
polynomial-time computable natural number $r_l$ such that
 the $r_l$-fold concatenation of $V_l$
solves the corresponding PSPACE problem, i.e.\
\[
V_l^{r_l} (\ket{x} \otimes \ket{y} \otimes \ket{00\dots 0}) =
|x\rangle \otimes |y \oplus f(x)\rangle  \otimes |00\dots 0\rangle\,,
\]
where $f$ is the characteristic function of $L$, i.e., $f(x)=1$ if
$x\in L$ and $f(x)=0$ otherwise. The vector $|x\rangle$ is the basis
state given by the binary word $x \in \{0,1\}^l$, the vector
$|y\rangle$ is the state of the output qubit and $|00\dots 0\rangle$ is
the initial state of $m_l-l-1$ ancilla qubits.
\end{theorem}
\proof{In order to construct the circuit $V_l$ corresponding to a
PSPACE problem we need to have an upper bound for the required
space. This is, for instance, the case for the PSPACE-complete 
problem QBF (Quantified-Boolean Formulas).
It can be solved within the space $O(l^2)$ where $l$
is the length of the input.
 This space bound determines
$m_l$, the number of qubits.

Let $M$ be a Turing machine that solves QBF within space $O(l^2)$. Now
we construct a quantum circuit $V_l$ that simulates the Turing machine
$M$ for input length $l$. 
Since the
computational steps of a quantum circuit are unitary (thus
reversible), we have to work with a reversible Turing machine $R$
instead of $M$ (the latter could be irreversible). 
Each application of the constructed circuit
simulates one or two steps of $R$.

Due to a result of Lange et al. (Theorem 3.3 in \cite{LMT:00}) it is
possible to simulate irreversible Turing machines by reversible ones
without increasing the necessary space too much. More precisely, they
give the simulation of a space-bounded Turing machine $M$ by a
reversible Turing machine $R$ operating on the same space.
In general, the reversible simulation by $R$ may
have an exponential time overhead. The running time overhead 
is not relevant here
because we can derive an upper bound on the 
running time of the reversible machine from the required 
number of qubits.
In the following we work with the reversible 
Turing machine $R$.

The fact that every Turing machine can be simulated efficiently by
circuits is standard \cite{Savage:00}. Here we need an explicit
construction converting the {\it reversible} Turing machine into a
circuit consisting of {\it reversible} gates.

The circuit acts on the following registers:
\begin{enumerate}
\item The register \texttt{head} encodes the internal state of the
Turing machine.
\item The register \texttt{tape\_index} stores the current location of
the head.
\item The register \texttt{ACC} is the accumulator (temporary
storage).
\item The register \texttt{tape} corresponds to a sufficiently large
region of the tape that is required for computation.  It consists of
cell $1$ to cell $N_l$ where $N_l$ is the space bound
corresponding to the input length $l$.
\end{enumerate}
Each step of the reversible Turing machine of Lange et al.\ is either
a moving or read-and-write transition\footnote{This separation is
useful in order to characterize reversibility of Turing machines
\cite{Bennett:89}.}. A moving transition has the form $p\rightarrow
(q,\pm 1)$. That means that in state $p$ the machine makes one step to
the right ($+1$) (respectively to the left ($-1$)) and changes into
state $q$ without reading or writing any tape cell. A read-and-write
transition has the form $(p,a)\rightarrow (q,b)$ meaning that in the
state $p$ the machine overwrites the symbol $a$ with the symbol $b$
and changes into state $q$ without moving the head. 

Furthermore, in our construction it is determined by the state of the
head whether the system performs a moving or a read-and-write
operation (and not by the state of the tape). In other words, the
state set $\hat{Q}$ of $R$ is the disjoint union of a set $Q$ of
read-and-write states, a set $Q^{\rightarrow}$ of right-moving states
and a set $Q^{\leftarrow}$ of left-moving states.

Due to reversibility of $R$ the moving transitions can be implemented
as a unitary transformation on the registers \texttt{head} and
\texttt{tape\_index}. A right-moving transition ($p\in
Q^{\rightarrow}$) translates as follows:
\begin{equation}\label{EQ1}
\ket{p}_{\texttt{head}}\otimes\ket{i}_{\texttt{tape\_index}}
\rightarrow
\ket{q}_{\texttt{head}} \otimes \ket{i+1}_{\texttt{tape\_index}}
\,.
\end{equation}
Analogously, a left-moving transition ($p\in Q^{\leftarrow}$)
translates as follows:
\begin{equation}\label{EQ2}
\ket{p}_{\texttt{head}}\otimes\ket{i}_{\texttt{tape\_index}}
\rightarrow
\ket{q}_{\texttt{head}} \otimes \ket{i-1}_{\texttt{tape\_index}}
\,.
\end{equation}
Note that the operations on the register \texttt{tape\_index} are
computed modulo $N$, where $N$ is the number of tape cells. Although
the Turing machine will never move to the right when the head is at
position $N$ and never to the left when it is at position $1$, this
definition guarantees that eqs.~(\ref{EQ1}) and (\ref{EQ2}) define
unitary operators.

These transformations can be realized efficiently as a unitary
transformation $U_{\rm moving}$ acting only on the registers
\texttt{head} and \texttt{tape\_index}.

Again, due to reversibility of $R$ the read-and-write transitions can
be realized by a unitary transformation. Let $(p,a)\rightarrow (q,b)$
be a read-and-write transition. There is a unitary transformation
$W_{\rm r/w}$ acting on the registers \texttt{head} and \texttt{ACC}
realizing
\[
\ket{p}_{\texttt{head}} \otimes \ket{a}_{\texttt{ACC}}
\rightarrow
\ket{q}_{\texttt{head}} \otimes \ket{b}_{\texttt{ACC}}
\,.
\]
We denote by SWAP(\texttt{ACC},\texttt{tape[i]}) the unitary operation
that swaps \texttt{ACC} and the $i$th cell of \texttt{tape}. We denote
by $\Lambda_i(U)$ the controlled operation that performs $U$ if and
only if \texttt{tape\_index} has the value $i$. Now we define
$U_{\rm r/w}$ as the concatenation
\[
\prod_{i=1}^N \Lambda_i\big(
{\rm SWAP}(\texttt{ACC},\texttt{tape[i]})\big)\,
W_{\rm r/w}\,
\prod_{i=1}^N \Lambda_i\big(
{\rm SWAP}(\texttt{ACC},\texttt{tape[i]})\big)\,.
\]
The transformation $U=U_{\rm r/w} U_{\rm moving}$ is the
transformation that corresponds to a moving and/or read-and-write
transition of the reversible Turing machine $R$. (Note that if a
read-write translation is followed by a moving transition then this $U$
performs both transitions.)

The constructed circuit $U$ does not satisfy all requirements of the
theorem. The first problem is that we do not know the running time of
the reversible Turing machine $R$. Consequently, we do not know how
many times we have to perform the elementary circuit $U$ to obtain the
solution. Even if we knew how many times we have to apply $U$, the
corresponding transformation would in general change the state
$\ket{x}$ and produce some garbage on the ancillas.

To circumvent the first problem we introduce some idle cycles to
guarantee that the running time is an efficiently computable function
$r_l$ of the input length $l$. The second problem is solved by
uncomputing the operations carried out during the computational steps
and the idle cycles.

In the step $r_l/2$ (proper computational steps and the idle cycles)
the solution is copied to the register \texttt{solution}. In the
following $r_l/2$ steps we uncompute the idle cycles and the
computational steps. The computational steps are uncomputed by
applying $U^\dagger$ corresponding to running $R$ backwards.

Now we construct the quantum circuit $V$ that circumvents both
problems as explained.  Note that we drop the index $l$ in the
following. The circuit operates on the registers \texttt{head},
\texttt{ACC}, \,\,\, \texttt{tape\_index},\texttt{tape} and the new
registers \texttt{solution}, \texttt{operation\_mode},
\texttt{idle\_counter}, and \texttt{counter} (See
Fig.~\ref{fig:whole_circuit}).

\begin{figure}
\centerline{\epsfxsize0.9\textwidth\epsfbox[10 20 390 490]{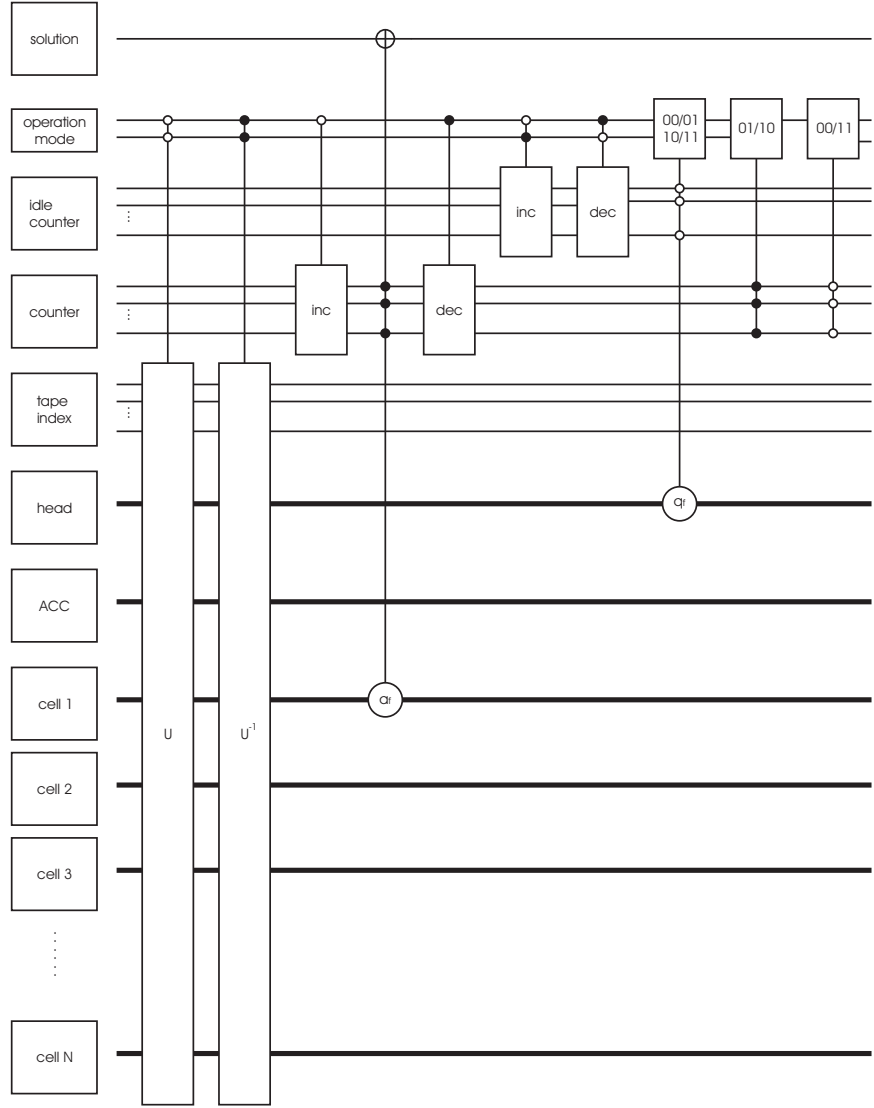}}
\caption{Quantum circuit satisfying the requirements of Theorem~1.
The gates INC and DEC increment and decrement the register counter and
idle counter, respectively. The gates $b_1 b_2/b_1' b_2'$ swaps the
state $\ket{b_1 b_2}$ and $\ket{b_1' b_2'}$. The gate $00/01\, 10/11$
is controlled by $q_f$ and the state of the idle counter. The symbol
$q_f$ represents all final states of the Turing machine $R$. The
symbol $a_f$ denote the solution $f(x)$. The bit-flip on the register
\texttt{solution} is controlled by $a_f$ and the state of the register
\texttt{counter}.}
\label{fig:whole_circuit}
\end{figure}

The register \texttt{operation\_mode} indicates whether the current
operation is $U$, idle cycle, reverse idle cycle, or $U^\dagger$.  
These $4$ subroutines of the whole circuit can be seen
in Fig.~\ref{4Phasen}.

\begin{figure}
\centerline{\epsfxsize1.0\textwidth\epsfbox{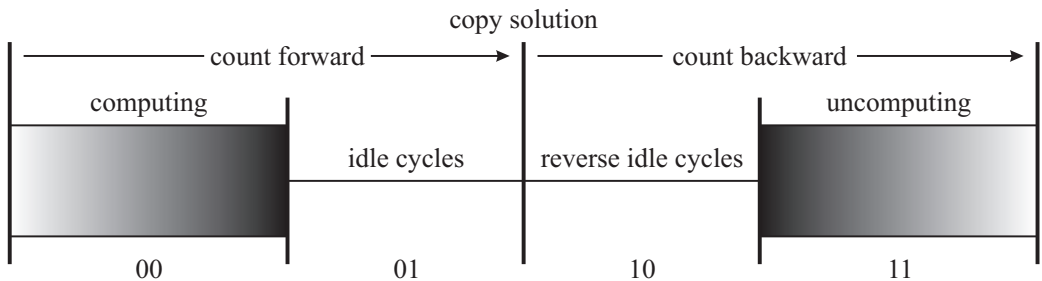}}
\caption{The $4$ subroutines of the circuit $V$. Simultaneously with the
subroutines the counter runs forward or backward.}
\label{4Phasen}
\end{figure}

The content of the register \texttt{counter} is incremented after each
application of $U$ or {\it idle cycle} and decremented after each
application of $U^\dagger$ or {\it reverse idle cycle}. Our
construction uses the following upper bound on the number of necessary
applications of $U$. Since the number of basis states of the register
that $U$ acts on is $2^m$ it does not make sense to have $r>2^m$.
Therefore the counter is incremented until the register has the result
$2^{m+1}-1$ in order to ensure that the number of required
applications of $U$ is exceeded. As soon as this number is reached the
result of the computation is copied to the register \texttt{solution},
i.e., the register is incremented by $1$ if and only if the answer is
``true''. From this moment on the \texttt{counter} and the
\texttt{idle\_counter} are decremented.  As soon as the
\texttt{idle\_counter} reaches $0$ the operation mode is changed such
that the concatenated application of $U^\dagger$ is started. After the
appropriate number of applications the initial state of all registers
are restored except from the register \texttt{solution} which is
incremented by $1$ if and only if the answer of the PSPACE problem is
``true''.

Explicitly, one has the following rules:
\begin{enumerate}
\item operation mode ${\bf 00}$: perform the circuit $U$, increment
\texttt{counter}
\item change operation mode ${\bf 00\rightarrow 01}$ if 
\texttt{idle\_counter} is $00\cdots 0$ and \texttt{head} is in a
final state
\item operation mode ${\bf 01}$: increment \texttt{counter} and
\texttt{idle\_counter}
\item increment \texttt{solution} if \texttt{operation\_mode} is $01$,
\texttt{counter} is $11\cdots 1$ and the first tape cell is in a state
indicating if the answer is true (we assume that this tape cell 
contains the result $f(x)$)
\item change operation mode ${\bf 01\rightarrow 10}$ if \texttt{counter} is
in $11\cdots 1$
\item operation mode ${\bf 10}$: decrement \texttt{counter} and
\texttt{idle\_counter}
\item change operation mode ${\bf 10\rightarrow 11}$ if
\texttt{idle\_counter} is in $00\cdots 0$ and \texttt{head} is in a
final state
\item operation mode ${\bf 11}$: perform the circuit $U^\dagger$,
decrement \texttt{counter} and \texttt{idle\_counter}
\item change operation mode ${\bf 11\rightarrow 00}$ if
\texttt{counter} is in $00\cdots 0$
\end{enumerate}
Note that the circuit $V$ has the following property: applied to the
initial state $|x\rangle |0\dots 0\rangle$ the orbit length is
$r=2(2^{m+1}-1)$ if the answer is ``false'' and $2 r$ whenever the
answer is ``true''.}

The dependence of the orbit length on the solution is essential in the
following section.

\section{Constructing the observable}
In this section we construct a family of observables $(A_l)$ in such a
way that the spectral properties of $A_l$ reflect the length of the
orbit $(V_l^j\ket{x}\otimes\ket{00\dots 0})_{j\in \N}$ for inputs of
length $l$. The idea to construct Hamiltonians corresponding to
quantum circuits already appeared in \cite{KR:03}. In this article,
the purpose was to show that a closed quantum (Hamiltonian) system can
in principle implement a circuit without any external control
operations. Similar constructions were also used in the context of
complexity theory in order to show that determining the spectrum of
physical Hamiltonians may be computationally hard
\cite{KitaevShen,KempeRegev}. However, their constructions deal with
quantum circuits of polynomial size.  The whole sequence of gates is
in some sense encoded into the Hamiltonian.  The solution of a NP or
QMA problem is then reflected in the least eigenvalue of the
Hamiltonian. The fact that the determination of the least eigenvalue
encompasses NP or QMA even if only inverse polynomial accuracy is
required is due to the polynomial length of the program. Here we have
typically an exponential number of applications and the solution of
the problem is therefore encoded in the ``hyperfine structure'' of the
spectrum.

Let $V$ be a quantum circuit as in Theorem~\ref{th:pspace} and $s$
be its size, i.e., the number of elementary two-qubit gates. We need a
register \texttt{clock} indicating which gate is applied. It consists
of $s_l$ qubits. The allowed states of the register \texttt{clock} 
are of the form
$
\ket{0\cdots 0 1 0 \cdots 0}
$
indicating which gate of $V$ is applied currently. 
We denote by $V_j$ the
elementary gates of $V$ (in contrast to the preceeding section
where the index denoted the input length).

We first define the {\em forward-time} operator
\begin{eqnarray*}
F & = &
V_1 \otimes \ket{1}_{2}\bra{0}_{2} \otimes \ket{0}_{1}\bra{1}_{1} + \\
& &
V_2 \otimes \ket{1}_{3}\bra{0}_{3} \otimes \ket{0}_{2}\bra{1}_{2} + \\
& &
\,\vdots \\
& &
V_{s} \otimes \ket{1}_{1}\bra{0}_{1} \otimes \ket{0}_{s}\bra{1}_{s}
\,.
\end{eqnarray*}
The operators $V_j$ operate on all registers of the preceding
section. The operators $|0\rangle_i \langle 1|_i$ and $|1\rangle_i
\langle 0|_i$ are annihilation and creation operators, respectively,
on the $i$th qubit of the \texttt{clock}.

We denote the linear span of the vectors
\[
F^j|\Psi_0\rangle \quad\mbox{for } j\in \N
\]
with $|\Psi_0\rangle := \ket{x}\otimes\ket{00\cdots 0}
\otimes\ket{100\cdots 0}$ as $\cO$. All states of this orbit are
orthogonal until one has recurrence to the initial state
$|\Psi_0\rangle$.  This can be seen as follows: If the register
\texttt{clock} is in an allowed state there is only one summand of $F$
that is relevant.  Its action on the \texttt{clock} is simple since it
moves the $1$ to the next qubit. Therefore it is clear that the first
$s-1$ states are orthogonal.  The whole circuit $V$ is a classical
logical operation which permutes basis states. Therefore the state
$F^s |\Psi_0\rangle$ is either orthogonal to $|\Psi_0\rangle$ or both
states coincide. Along the same line we argue that all states of the
orbit are orthogonal until a state coincides with the initial
state. Hence $F$ acts as a cyclic shift on $\cO$.

The dimension of $\cO$ is $2 s r$ if $f(x)=1$ and $s r$ if
$f(x)=0$. We
denote the dimension by $d$.

Let $\omega$ be a primitive complex $d$-th root of unity. The
eigenvalues of $F$ restricted to $\cO$ are
\[
\omega^0,\omega^1,\omega_2,\ldots,\omega^{d-1}\,.
\]
Furthermore, the initial state vector $\ket{\Psi_0}$ is a
superposition of all eigenvectors of $F$ restricted to $\cO$ with
equal weights. All this follows from properties of the cyclic shift
operator.

The backward-time operator is defined as the adjoint of $F$.  The
observable $A$  is defined as the sum of the forward and backward time
operators, i.e., $A:=(F+F^\dagger)/2$. 
It is $4$-local since each
$V_j$ is $2$-local and is coupled to a $2$-local propagator. The
dynamics of the \texttt{clock} may be interpreted as a propagation of
a spin-wave. Note that the idea of our construction is not to
implement the quantum circuit $V$ by the autonomous time evolution
$\exp(-i A t)$. 
The aim is rather to obtain an observable such that
its spectral properties correspond to the orbit length of the circuit.

Since $F$ and $F^\dagger$ commute on $\cO$ the eigenvalues of $A$
restricted to $\cO$ are $(\omega^j+\bar{\omega}^j)/2=\cos(2\pi j/d)$.
The non-real eigenvalues are $2$-fold degenerated. Only
the eigenvalues $1$ and $-1$ have multiplicity $1$.
In a hypothetical energy measurement
applied to the initial state vector $\ket{\Psi_0}$
 one would obtain all
$2$-fold degenerated eigenvalues  with probability $2/d$ each and the
non-degenerated eigenvalues with probability $1/d$.
Note that only the first case is relevant for large $d$ 
 since there are at most two
non-degenerated values.

Note that $d$ depends on the solution of the PSPACE problem.
Explicitly, the possible measurement results are

\begin{enumerate}
\item either
\[
\cos(2\pi j/ (2 s\, r))\,,\quad j=0,\ldots,2s\, r-1
\]
\item or
\[
\cos(2\pi j/ (s\, r))\,,\quad j=0,\ldots,s \,r-1
\]
\end{enumerate}
depending on whether $f(x)=1$ or $f(x)=0$. 

Note that a perfect energy measurement can distinguish between the two
cases even after few samples: after applying the function
``$\arccos$'' we obtain values with distance $2\pi/d$ and all values
occur with equal probability (if the non-degenerated values are
neglected). Then it is easy to distinguish between the two cases
$d=r\,s$ and $d=2\,r\,s$.

Now we examine what accuracy is sufficient to distinguish between the
two cases. For doing so, we will restrict our attention to those
measurement values which are between $1/\sqrt{2}$ and
$-1/\sqrt{2}$. This means that half of the measurement outcomes have
to be ignored because the probability to obtain an outcome in this
interval is about $1/2$. These values correspond to angles in the
interval $[\pi/4,3\pi/4]$ and $[5\pi/4,7\pi/4]$.

In the following we assume that we have obtained a measurement value
in this interval. For each outcome $E$ we chose $j$ such that
$|\arccos(E)-2\pi j/ (2 r s)|$ is minimal. If $f(x)=1$ then the
probability of obtaining an odd value for $j$ is at least
$1/2\,\cdot\,3/4=3/8$. If $f(x)=0$ then the probability of obtaining
an even value for $j$ is at least $1\,\cdot\,3/4=3/4$. Therefore, the
probability of odd value is at most $1/4$. This difference in
probability allows to distinguish between the two cases. It is obvious
that the error probability decreases exponentially with the number of
measurements.

Note that the observable $A$ has spectral gaps that are considerably
smaller than the required accuracy. This can already be seen if we
consider the $A$-invariant subspace $\cO$. The distance of the largest
eigenvalue $1$ and the second largest eigenvalue $\cos(2\pi/d)$ of $H$
is approximatively given by $(2\pi/d)^2$ since the derivative of the
cosinus function at $0$ is $0$.

Note that the required accuracy is directly connected with an upper
bound on the running time $T$. In our setting the running time is the
number $r$ of necessary applications of the circuit $V$ times the
number $s$ of gates of $V$.

In the construction of the preceding section we obtained the upper
bound on $rs$ from the required space.  More generally, whenever we
know that $r$ applications of $V$ are sufficient we need a measurement
with accuracy of the order $1/ (r\, s)$ to determine the solution of
the PSPACE problem.

This discussion proves the following theorem:

\begin{theorem}[Measurement precision vs. running time]${}$\\
Let $\{A_l\}$ be a family of $4$-local observables corresponding to
the family $\{V_l\}$ of quantum circuits in Theorem~1.  Then every
measurement in the sense of Definition~\ref{def:measure}
could be used to solve PSPACE problems in polynomial time whenever the
accuracy is of the order of the spectral gaps of $H$. It is even
sufficient to have an measurement error $1/T_l$, where $T_l$ is the
running time of the algorithm based on the circuit $V_l$.
\end{theorem}

\section{Conclusions}
We have shown that every apparatus which implements {\it precise}
measurements of $4$-local $n$-qubit observables would solve PSPACE
problems.  This conclusion does only hold for exponentially small
errors of the measurement.  On the other hand, we have argued that
algorithms which measures with inverse-polynomial accuracy can be
implemented efficiently.  Provided that PSPACE problems cannot be
solved efficiently, i.e., PSPACE $\neq$ BQP, the complexity of
measurements depend on the required accuracy. The statement that
exponential accuracy has stronger computational power is also
well-known in classical analog computational models
\cite{Hartmanis,Vergis}.

One may ask why one should try to measure general $4$-local observables.
A possible motivation to develop a complexity theory of measurements is
that some proposals for quantum algorithms use 
joint observables on the quantum register \cite{Ettinger}.
 
Another motivation is that one is interested in measurements for
physically relevant joint observables like energy. The reader may
object that the specific interactions constructed in this paper are
rather unphysical for several reasons:

\begin{enumerate}

\item Most interactions in nature are pair-interactions and not $4$-local.

\item Our construction uses long-range interactions among distant
qubit quadruples. 

\item Interactions in natural many-particle systems have typically  
high symmetry. For instance, the interactions in solid states 
respect the translational invariance of the lattice.

\item There exist only a few fundamental interactions in physics.

\end{enumerate}

We have already argued that pair-interactions between particles may
correspond to $k$-local terms if some qubits encode the physical state
of one particle. This refutes the first objection.

We conjecture that the solution of PSPACE problems would even be
possible if the class of observables was restricted to those which
appear as Hamiltonians of real many-particle systems. This conjecture
is supported by the following ideas:

Quantum cellular automatons like the Hamiltonian dynamical system
constructed in \cite{Margolus:90} are also computationally
universal. Hamiltonians for those types of cellular automata have the
property that every cell interacts only with some cells in its
neighborhood. By merging some cells together to one cell we can always
obtain a Hamiltonian with pair-interactions among qudits. This seems
to indicate that neither the symmetry nor locality assumptions on the
interactions prevents the Hamiltonian from corresponding to
computationally universal networks.  Due to the fact that computers
exist it is clear that the structure of the fundamental interactions
is general enough to allow universal systems.  Therefore we guess that
spectral properties of more realistic Hamiltonians encode PSPACE
problems in a similar way as in our paper.

\subsection*{Acknowledgements}
Thanks to J\"{o}rn M\"{u}ller-Quade and Markus Grassl for helpful
discussions. 
This work has been supported by grants of the {\it Landesstiftung
Baden-W\"{u}rttemberg} (project ``Kontinuierliche Modelle der
Quanteninformationsverarbeitung'').


\end{document}